\documentclass[10pt,twocolumn,letterpaper]{article}

\usepackage{cvpr}              

\usepackage{graphicx}
\usepackage{amsmath}
\usepackage{amssymb}
\usepackage{booktabs}

\usepackage{multirow}
\usepackage{caption}
\usepackage{float}
\usepackage{url}
\usepackage{enumitem}
\usepackage[ruled]{algorithm2e}
\usepackage[accsupp]{axessibility}

\usepackage[pagebackref,breaklinks,colorlinks]{hyperref}

\usepackage[capitalize]{cleveref}
\crefname{section}{Sec.}{Secs.}
\Crefname{section}{Section}{Sections}
\Crefname{table}{Table}{Tables}
\crefname{table}{Tab.}{Tabs.}

\def\cvprPaperID{1134} 
\def\confName{CVPR}
\def\confYear{2022}

\newcommand*{\affaddr}[1]{#1} 
\newcommand*{\affmark}[1][*]{\textsuperscript{#1}}

\begin{document}

\title{ Blind2Unblind: Self-Supervised Image Denoising with Visible Blind Spots}

\author{Zejin Wang\affmark[1,2]\qquad Jiazheng Liu\affmark[1,3]\qquad Guoqing Li\affmark[1]\qquad Hua Han\affmark[1,3,]\thanks{Corresponding author}\\
 \affaddr{\normalsize\affmark[1]National Laboratory of Pattern Recognition, Institute of Automation, Chinese Academy of Sciences}\\
 \affaddr{\normalsize\affmark[2]School of Artificial Intelligence, University of Chinese Academy of Sciences}\\
 \affaddr{\normalsize\affmark[3]School of Future Technology, University of Chinese Academy of Sciences}\\
}
\maketitle

\begin{abstract}
   Real noisy-clean pairs on a large scale are costly and difficult to obtain. Meanwhile, supervised denoisers trained on synthetic data perform poorly in practice. Self-supervised denoisers, which learn only from single noisy images, solve the data collection problem. However, self-supervised denoising methods, especially blindspot-driven ones, suffer sizable information loss during input or network design. The absence of valuable information dramatically reduces the upper bound of denoising performance. In this paper, we propose a simple yet efficient approach called Blind2Unblind to overcome the information loss in blindspot-driven denoising methods. First, we introduce a global-aware mask mapper that enables global perception and accelerates training. The mask mapper samples all pixels at blind spots on denoised volumes and maps them to the same channel, allowing the loss function to optimize all blind spots at once. Second, we propose a re-visible loss to train the denoising network and make blind spots visible. The denoiser can learn directly from raw noise images without losing information or being trapped in identity mapping. We also theoretically analyze the convergence of the re-visible loss. Extensive experiments on synthetic and real-world datasets demonstrate the superior performance of our approach compared to previous work. Code is available at \href{https://github.com/demonsjin/Blind2Unblind}{https://github.com/demonsjin/Blind2Unblind}.
\end{abstract}

\section{Introduction}
\label{sec:intro}
Image denoising, an essential task of low-level image processing, aims to remove noise and restore a clean image. In vision applications, the quality of denoising significantly affects the performance of downstream tasks, such as super-resolution~\cite{huang2019simultaneous}, semantic segmentation~\cite{liu2020connecting}, and object detection~\cite{shijila2019simultaneous}. In addition, the denoiser can significantly improve the quality of images captured by mobile phones and other devices, reflecting a broad demand in imaging fields.

With the development of neural networks, learning-based denoisers~\cite{ronneberger2015u,zhang2017beyond,zhang2018ffdnet,guo2019toward,anwar2019real,yue2019variational,zamir2020learning,chang2020spatial} have recently shown superior performance than traditional methods~\cite{dabov2007image,buades2005non,dabov2007color,gu2014weighted}. However, supervised denoisers, \eg, U-Net~\cite{ronneberger2015u}, DnCNN~\cite{zhang2017beyond}, FFDNet~\cite{zhang2018ffdnet}, RIDNet~\cite{anwar2019real}, SANet~\cite{chang2020spatial}, rely on numerous noisy-clean pairs, which are costly and hard to collect. The performance of denoisers drops dramatically once processing unknown noise patterns. Lehtinen \etal~\cite{lehtinen2018noise2noise} then propose to recover clean signals directly from corrupted image pairs. Using corrupted pairs reduces the difficulty of data collection but remains challenging for dynamic scenes with deformations and image quality variations.

To alleviate the above limitations, self-supervised denoising~\cite{ulyanov2018deep,batson2019noise2self,krull2019noise2void,wu2020unpaired,laine2019high,huang2021neighbor2neighbor,pang2021recorrupted} that learns from a single noisy image has attracted much interest from researchers. Ulyanov \etal~\cite{ulyanov2018deep} learn deep prior only from a single noisy image. Namely, each degraded image has to be trained from scratch. Manual masking, \eg, Noise2Self~\cite{batson2019noise2self}, Noise2Void~\cite{krull2019noise2void}, avoids custom denoising for each image. Since blind spots on the inputs occupy a large area, the receptive field of predicted pixels loses much valuable context, resulting in poor performance. Moreover, optimizing partial pixels in each iteration causes slow convergence. Laine \etal~\cite{laine2019high} design a blind spot network to bound the receptive field in four directions instead of manual masking. Masked convolution accelerates training and increases the receptive field to all areas except the blind spot. Similarly, the dilated blindspot network~\cite{wu2020unpaired} sets blindspots on the receptive field without masking the inputs. Regardless of masked input or blind-spot networks, lower accuracy limits practical applications. Bayesian estimation~\cite{laine2019high,wu2020unpaired,krull2020probabilistic} is used for explicit noise modeling as post-processing. However, noise modeling performs poorly on real-world data with complex patterns. Some works~\cite{xu2020noisy,moran2020noisier2noise} perform noise reduction for noisier-noise pairs even though the additional noise increases the information loss and requires that the extra noise has the same distribution as the original one. Subsequently, Pang \etal~\cite{pang2021recorrupted} develop a data augmentation technique with the known noise level to address the overfitting caused by the absence of truth images. Recently, Huang \etal~\cite{huang2021neighbor2neighbor} propose to train the network with training pairs sub-sampled from the same noisy image. However, using sub-sampling pairs for supervision lead to over smoothing as neighboring pixels are approximated.

In this paper, we propose Blind2Unblind, a novel self-supervised denoising framework that overcomes the above limitations. Our framework consists of a global-aware mask mapper based on mask-driven sampling and a training strategy without blind spots based on re-visible loss. Specifically, we divide each noisy image into blocks and set specific pixels in each block as blind spots, so that we can obtain a global masked volume as input, which consists of a set of images with order masks. Then, the volume with global masks is fed into the network in the same batch. The global mapper samples denoised volumes at blind spots and projects them onto the same plane to generate denoised images. The operation speeds up training, enables global optimization, and allows the application of re-visible loss. However, masked images result in a sizable loss of valuable information, severely reducing the upper bound of denoising performance. Therefore, we consider learning from raw noisy images without masks and relief from identity mapping. Furthermore, the intermediate medium of gradient update must be introduced since raw noisy images cannot participate in backpropagation during training. We assume that masked images serve as a medium and propose a re-visible loss to enable the transition from blind-spot denoising to non-blind denoising. The proposed self-supervised denoising framework, which does not involve noise model prior or the removal of valuable information, shows surprising performance. Moreover, advanced models can be applied to our proposed method.

The contributions of our work are as follows:
\begin{enumerate}[itemsep=2pt,topsep=0pt,parsep=0pt]
\item We propose a novel self-supervised denoising framework that makes blind spots visible, without sub-sample, noise model priors and identity mapping.
\item We provide a theoretical analysis of re-visible loss and present its upper and lower bounds on convergence.
\item Our approach shows superior performance compared with state-of-the-art methods, especially on real-world datasets with complex noise patterns.
\end{enumerate}

\section{Related Work}
\label{sec:related}

\subsection{Non-Learning Image Denoising}
Non-learning denoising methods such as BM3D~\cite{dabov2007image}, CBM3D~\cite{dabov2007color}, WNNM~\cite{gu2014weighted}, and NLM~\cite{buades2005non} usually iteratively perform custom denoising for each noisy image. However, since the denoising procedure is iteratively adjusted and the hyperparameters are limited, these methods suffer from long inference time and poor performance.

\subsection{Supervised Image Denoising}
Recently, many supervised approaches for image denoising have been developed. Zhang \etal~\cite{zhang2017beyond}, the pioneer in deep image denoising, first proposes DnCNN to process unknown noise levels using noisy-clean pairs as supervision. Then, U-Net~\cite{ronneberger2015u} becomes a common denoiser based on multi-scale features. After that, more advanced denoisers~\cite{zhang2018ffdnet,guo2019toward,anwar2019real,yue2019variational,zamir2020learning,chang2020spatial} are proposed to improve denoising performance under supervision. However, collecting noise-clean pairs remains challenging in practice since the cost of acquisition, deformation, and contrast variations. Moreover, supervision with a strong prior often performs poorly when noise patterns are unknown.

\subsection{Self-Supervised Image Denoising}
Noise2Noise~\cite{lehtinen2018noise2noise} uses noisy-noisy pairs for training, which reduces the difficulty of data collection. Then, Noise2Self~\cite{batson2019noise2self} and Noise2Void~\cite{krull2019noise2void} propose masked schemes for denoising on individual noisy images. Since some areas on noisy images are blind, the accuracy of denoising is low.  Laine19~\cite{laine2019high}, DBSN~\cite{wu2020unpaired}, and Probabilistic Noise2Void~\cite{krull2020probabilistic} transfer the masking procedure to the feature level for larger receptive fields, which reduces valuable information loss. Moreover, noise model priors are also introduced as post-processing to improve performance. Self2Self~\cite{quan2020self2self} trains a dropout denoiser on the pair generated by the Bernoulli sampler and averages the predictions of multiple instances. Noisy-As-Clean~\cite{xu2020noisy} and Noisier2Noise~\cite{moran2020noisier2noise} introduce added noise to train the denoiser and require a known noise distribution, limiting their use in practice. Similarly, R2R~\cite{pang2021recorrupted} also corrupts noisy images with known noise levels to obtain training pairs. Recently, Neighbor2Neighbor (NBR2NBR)~\cite{huang2021neighbor2neighbor} obtains the noise pair for training by sub-sampling the noise image. However, approximating neighbor pixels leads to over smoothing, and sub-sampling destroys the structural continuity.

\section{Theoretical Framework}
\label{sec:teo}

\subsection{Motivation}
\label{sec:mov}
Classical blindspot denoising methods, \eg, masked inputs~\cite{krull2019noise2void,batson2019noise2self,krull2020probabilistic} and blindspot networks~\cite{laine2019high,wu2020unpaired}, use an artificial masking scheme to form blind-noisy pairs. However, since the mask prior is suboptimal and lossy, their performance is severely limited. Intuitively, denoising a raw noisy image without blind spots can solve the performance degradation. Given a single raw noisy image, we assume that the model can perform denoising without losing valuable information. The only thing is to teach the model how to learn and what to learn. Since the model now has to learn denoising from the raw noisy image without blind spots, eliminating identity mapping is critical. We consider that using the masked input as a medium for updating the gradient prevents identity mapping. The challenge now remains to design a novel loss that associates blind spot denoising with raw noisy image denoising.

\subsection{Noise2Void Revisit}
\label{sec:n2v}
Noise2Void~\cite{krull2019noise2void} is a self-supervised denoising method using a training scheme that does not require noisy image pairs or clean target images. The approach only requires a single noisy image and then applies a blind-spot masking scheme to generate blind-noisy pairs for training. Given the noisy observation $\mathbf{y}$ of the ground truth $\mathbf{x}$, Noise2Void aims to minimize the following empirical risk:
\begin{equation}
\label{eq:1}
\mathop{\arg\min}\limits_{\theta} \mathbb{E}_{\mathbf{y}}\Vert f_{\theta}(\mathbf{y}_{{\rm RF}(i)})-\mathbf{y}_i\Vert_2^2,
\end{equation}
where $f_{\theta}(\cdot)$ denotes the denoising model with parameter $\theta$, $\mathbf{y}_{{\rm RF}(i)}$ is a patch around pixel $i$, $\mathbf{y}_i$ means a single pixel $i$ located at the patch center. This method assumes that noise is context-independent while signal relies on surrounding pixels. The blind spot architecture takes advantage of the above properties and thus relief itself from identity mapping. 

\subsection{Re-Visible without Identity Mapping}
\label{sec:rev}
The blind spot scheme~\cite{krull2019noise2void,batson2019noise2self} can use less information for its prediction. Therefore, its accuracy is lower than that of the normal network. The challenge here is how to convert the invisible blind spots into visible ones. In this way, we can use the structure of blind spots for self-supervised denoising and then use all the information to improve its performance. First, we see the loss in the multi-task form of denoising as follows:
\begin{equation}
\label{eq:2}
\mathop{\arg\min}\limits_{\theta} \mathbb{E}_{\mathbf{y}}\Vert h(f_{\theta}(\mathbf{\Omega_{y}}))-\mathbf{y}\Vert_2^2+\lambda\cdot \Vert f_{\theta}(\mathbf{y})-\mathbf{y}\Vert_2^2,
\end{equation}
note that $\mathbf{\Omega_{y}}$ denotes a noisy masked volume in which the contained blind spots are exactly at all positions of $\mathbf{y}$, $h(\cdot)$ is a global-aware mask mapper that samples the denoised pixels where the blind spots are located, and $\lambda$ is a constant. Using~\cref{eq:2} as the objective function for training learns the identity. That is, $f_{\theta}(\mathbf{y})$ cannot explicitly participate in backpropagation. We hope this element can implicitly participate in the gradient update and realize the transition from blind to non-blind. We reformulate~\cref{eq:2} in the 1-norm form:
\begin{equation}
\label{eq:3}
\mathop{\arg\min}\limits_{\theta} \mathbb{E}_{\mathbf{y}}\Vert h(f_{\theta}(\mathbf{\Omega_{y}}))-\mathbf{y}\Vert_1+\lambda\cdot \Vert \hat{f}_{\theta}(\mathbf{y})-\mathbf{y}\Vert_1,
\end{equation}
where $\hat{f}_{\theta}(\cdot)$ means that they do not contribute to updating the gradient. That is, $\hat{f}_{\theta}(\mathbf{y})$ can be considered as an ordinary constant. It only remains to design a new objective function and the derivative of $f_{\theta}(\mathbf{\Omega_{\mathbf{y}}})$ should contain $\hat{f}_{\theta}(\mathbf{y})$ to satisfy the non-blind requirement. We assume that the quadratic operation on ~\cref{eq:3} can achieve the implicit goal of re-visible and satisfy the average arithmetic form of the optimal solution for the denoising task, which then becomes:
\begin{equation}
\label{eq:4}
\mathop{\arg\min}\limits_{\theta} \mathbb{E}_{\mathbf{y}}\Vert\vert h(f_{\theta}(\mathbf{\Omega_{y}}))-\mathbf{y}\vert+\lambda\cdot \vert \hat{f}_{\theta}(\mathbf{y})-\mathbf{y}\vert\Vert^2_2,
\end{equation}
where $\vert\cdot\vert$ denotes the absolute value of each element in the vector. Direct application of~\cref{eq:4} as the objective function is not appropriate, since the symbol of different terms reflects opposite optimization objectives. Specifically, we extend~\cref{eq:4} as follows:
\begin{equation}
\label{eq:5}
\begin{split}
\mathcal{T}(\mathbf{y})&=\Vert\vert h(f_{\theta}(\mathbf{\Omega_{\mathbf{y}}}))-\mathbf{y}\vert+\lambda\cdot \vert \hat{f}_{\theta}(\mathbf{y})-\mathbf{y}\vert\Vert^2_2 \\
&= \Vert h(f_{\theta}(\mathbf{\Omega_{\mathbf{y}}}))-\mathbf{y}\Vert^2_2+\lambda^2\Vert \hat{f}_{\theta}(\mathbf{y})-\mathbf{y}\Vert^2_2 \\
&\quad+2\lambda\Vert(h(f_{\theta}(\mathbf{\Omega_{\mathbf{y}}}))-\mathbf{y})\odot(\hat{f}_{\theta}(\mathbf{y})-\mathbf{y}) \Vert_1.
\end{split}
\end{equation}
Let $d_1=h(f_{\theta}(\mathbf{\Omega_{\mathbf{y}}}))-\mathbf{y},d_2=\hat{f}_{\theta}(\mathbf{y})-\mathbf{y},cond=(h(f_{\theta}(\mathbf{\Omega_{\mathbf{y}}}))-\mathbf{y})\odot(\hat{f}_{\theta}(\mathbf{y})-\mathbf{y})$. Here, we divide the objective function $\mathcal{T}(\mathbf{y})$ into two cases:\\
i) If $cond\geqslant0$, we have 
\begin{equation}
\label{eq:6}
\mathcal{T}(\mathbf{y})=\Vert h(f_{\theta}(\mathbf{\Omega_{\mathbf{y}}}))+\lambda\hat{f}_{\theta}(\mathbf{y})-(\lambda+1)\mathbf{y}\Vert^2_2;
\end{equation}
ii) Similarly, when $cond<0$, it holds that 
\begin{equation}
\label{eq:7}
\mathcal{T}(\mathbf{y})=\Vert\hat{f}_{\theta}(\mathbf{y})-h(f_{\theta}(\mathbf{\Omega_{\mathbf{y}}}))+(\lambda-1)(\hat{f}_{\theta}(\mathbf{y})-\mathbf{y})\Vert^2_2.
\end{equation}
\begin{figure*}[ht]
\centering
\setlength{\abovecaptionskip}{0.cm} 
\includegraphics[width=\textwidth]{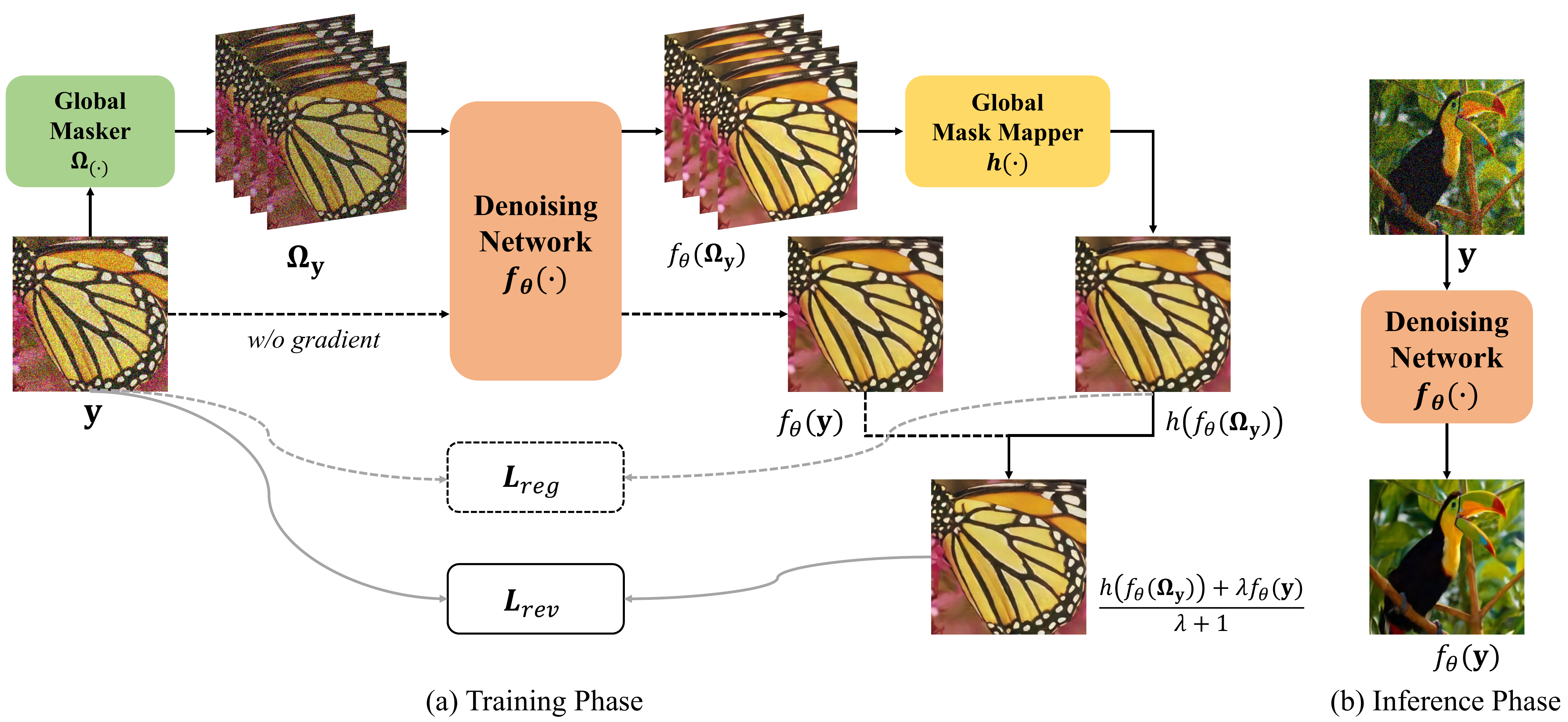}
   \caption{Overview of our proposed Blind2Unblind framework. (a) Overall training process. The global masker $\mathbf{\Omega_{(\cdot)}}$ creates a masked volume by adding blind spots to a noisy image $\mathbf{y}$. Then, the global-aware mask mapper samples the denoised volume to obtain $h(f_{\theta}(\mathbf{\Omega_{\mathbf{y}}}))$. Meanwhile, the denoiser $f_{\theta}(\cdot)$ takes $\mathbf{y}$ as input and produces the denoised result $f_{\theta}(\mathbf{y})$. The re-visible loss realizes the transition from the blind to visible with the invisible term $h(f_{\theta}(\mathbf{\Omega_{\mathbf{y}}}))$ as a medium. Moreover, the regular term is used to stabilize the training phase. (b) Inference using the trained denoising model. The denoising network generates denoised images directly from noisy images $\mathbf{y}$ without additional operation.}
\label{fig:overview}
\vspace{-0.4cm}
\end{figure*}
Note that when $cond<0$, the learning objective of $\hat{f}_{\theta}(\mathbf{y})$ becomes $h(f_{\theta}(\mathbf{\Omega_{\mathbf{y}}}))$, instead of $\mathbf{y}$ as~\cref{eq:6}. Considering the optimal denoiser $f^{*}_{\theta}$ that is trained to convergence, then $\hat{f}^{*}_{\theta}(\mathbf{y})$ should be approximate to $h(f^{*}_{\theta}(\mathbf{\Omega_{\mathbf{y}}}))$ in~\cref{eq:7}. However, the masking operation $\mathbf{\Omega_{(\cdot)}}$ causes information loss, resulting in low denoising upper bound for $h(f_{\theta}(\mathbf{\Omega_{\mathbf{y}}}))$. Besides, $\hat{f}_{\theta}(\mathbf{y})-h(f_{\theta}(\mathbf{\Omega_{\mathbf{y}}}))$ in~\cref{eq:7} indicates that $\hat{f}_{\theta}(\mathbf{y})$ should approximate $h(f_{\theta}(\mathbf{\Omega_{\mathbf{y}}}))$, which in turn suppresses the performance of $\hat{f}_{\theta}(\mathbf{y})$. Combining the above analysis, the final re-visible loss can be formulated as:
\begin{equation}
\label{eq:8}
\mathop{\arg\min}\limits_{\theta} \mathbb{E}_{\mathbf{y}}\Vert h(f_{\theta}(\mathbf{\Omega_{\mathbf{y}}}))+\lambda\hat{f}_{\theta}(\mathbf{y})-(\lambda+1)\mathbf{y}\Vert^2_2.
\end{equation}
When the denoiser converges to $f^{*}_{\theta}$, the following holds with the optimal solution $\tilde{\mathbf{x}}$ of~\cref{eq:8}:
\begin{equation}
\label{eq:9}
\tilde{\mathbf{x}} = \frac{h(f^{*}_{\theta}(\mathbf{\Omega_{\mathbf{y}}}))+\lambda\hat{f}^{*}_{\theta}(\mathbf{y})}{\lambda+1}.
\end{equation}
We assume that $h(f^{*}_{\theta}(\mathbf{\Omega_{\mathbf{y}}}))=\mathbf{x}+\varepsilon_1$ and $\hat{f}^{*}_{\theta}(\mathbf{y})=\mathbf{x}+\varepsilon_2$. Empirically, $\Vert \varepsilon_1\Vert_1 > \Vert \varepsilon_2\Vert_1$. With~\cref{eq:4,eq:8}, the upper and lower bounds of $\tilde{\mathbf{x}}$ are respectively $\hat{f}^{*}_{\theta}(\mathbf{y})$ and $h(f^{*}_{\theta}(\mathbf{\Omega_{\mathbf{y}}}))$. Namely, $h(f^{*}_{\theta}(\mathbf{\Omega_{\mathbf{y}}}))\leq\tilde{\mathbf{x}}\leq\hat{f}^{*}_{\theta}(\mathbf{y})$. Therefore, we can consider the generation of denoised images $\tilde{\mathbf{x}}$ only from noisy original images $\mathbf{y}$ during inference. Given $\mathbf{y}$, it holds that $\mathop{lim}\limits_{\lambda\rightarrow+\infty}\tilde{\mathbf{x}}=\hat{f}^{*}_{\theta}(\mathbf{y})$.

\subsection{Stabilize Transition Procedure}
\label{sec:stable}
As described in~\Cref{sec:rev}, the blind part $|h(f_{\theta}(\mathbf{\Omega_{\mathbf{y}}}))-\mathbf{y}|$ is used as a transition to optimize the re-visible one $\lambda\cdot|\hat{f}_{\theta}(\mathbf{y})-\mathbf{y}|$. Therefore, the cumulative error of the blind part can also affect the non-blind part. However, the pure re-visible loss lacks the separate constraint just for the transition term, which aggravates the instability during the training phase. We consider adding an extra constraint to correct $f_{\theta}(\mathbf{\Omega_{\mathbf{y}}})$, forcing the blind part to be zero. Thus, based on~\cref{eq:8}, we have the following constrained optimization problem:
\begin{equation}
\label{eq:10}
\begin{split}
&\mathop{\min}\limits_{\theta} \mathbb{E}_{\mathbf{y}}\Vert h(f_{\theta}(\mathbf{\Omega_{\mathbf{y}}}))+\lambda\hat{f}_{\theta}(\mathbf{y})-(\lambda+1)\mathbf{y}\Vert^2_2,\\
&\mbox{s.t.}\quad \Vert h(f_{\theta}(\mathbf{\Omega_{\mathbf{y}}}))-\mathbf{y}\Vert^2_2=0.
\end{split}
\end{equation}
We further reformulate it as the following optimization problem with a regularization term:
\begin{equation}
\label{eq:11}
\begin{split}
&\quad\mathop{\min}\limits_{\theta} \mathbb{E}_{\mathbf{y}}\Vert h(f_{\theta}(\mathbf{\Omega_{\mathbf{y}}}))+\lambda\hat{f}_{\theta}(\mathbf{y})-(\lambda+1)\mathbf{y}\Vert^2_2\\
&+\eta\cdot\Vert h(f_{\theta}(\mathbf{\Omega_{\mathbf{y}}}))-\mathbf{y}\Vert^2_2=0.
\end{split}
\end{equation}

\section{Main Method}
\label{sec:met}
Based on the theoretical analysis in~\Cref{sec:teo}, we propose Blind2Unblind, a novel deep self-supervised denoising framework learning from a single observation of noisy images. The framework consists of two main components: global-aware mask mapper and re-visible loss. The mask mapper generates masked volumes of noisy images as $\mathbf{\Omega}_{\mathbf{y}}$ and then samples denoised pixels in which there are blind spots to form the final denoised results. Moreover, we introduce a regularized re-visible loss to convert the blind spots into visible ones and overcome the challenge of having less information based on blind spots. An overview of our proposed Blind2Unblind framework can be found in~\Cref{fig:overview}.
\subsection{Global-Aware Mask Mapper}
\begin{figure}[ht]
\centering
\vspace{-0.32cm}
\setlength{\abovecaptionskip}{0.cm} 
\includegraphics[width=.47\textwidth]{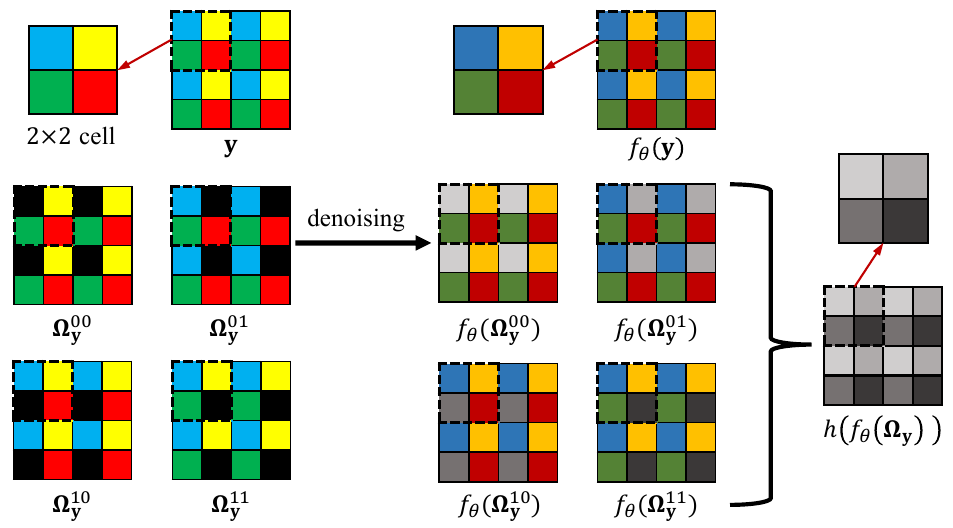}
   \caption{Details of a global-aware mask mapper. Taking a $2\times2$ cell $\mathbf{y}$ as an example, the global masker $\mathbf{\Omega_{(\cdot)}}$ hides four spots in $\mathbf{y}$ to create a global masked volumes $\mathbf{\Omega_{\mathbf{y}}}$ consisting of four masked cells ${\mathbf{\Omega}^{ij}_{\mathbf{y}}},i,j\in\{0,1\}$. The mask mapper $h(\cdot)$ samples the denoised volumes $f_{\theta}(\mathbf{\Omega_y})$ in which there are blind spots. The final denoised cell $h(f_{\theta}(\mathbf{\Omega_y}))$ is formed by the mapper based on the sampled locations and pixel values.}
\label{fig:mapper}
\vspace{-0.8cm}
\end{figure}

Manual masking methods~\cite{krull2019noise2void,batson2019noise2self} hide part of pixels for training, and objective function focuses only on masked regions, leading to a decrease in accuracy and slow convergence. The mask mapper performs global denoising on blind spots. This mechanism constrains all pixels, promotes information exchange across whole masked regions, and increases noise reduction accuracy. In addition, our mask mapper also speeds up the training of manual masking schemes.
The workflow of masked volumes generation and global mapping using mask mapper is shown in~\Cref{fig:mapper}. Given a noisy image $\mathbf{y}$ with width $W$ and height $H$, details of the global-aware mask mapper $h\left(f_{\theta}\left(\mathbf{{\Omega}_{y}}\right)\right)$ are described as follows:
\begin{enumerate}[itemsep=2pt,topsep=0pt,parsep=0pt]
\item First, the noisy image $\mathbf{y}$ is divided into $\lfloor{W/s}\rfloor\times\lfloor{H/s}\rfloor$ cells, each of size $s\times s$. For simplicity, we set $s=2$.
\item The pixels in the $i$-th row and $j$-th column of each cell are masked and filled with black for illustration. Thus, the masked image $\mathbf{\Omega}_{\mathbf{y}}^{ij}$ consists of $\lfloor{W/s}\rfloor\times\lfloor{H/s}\rfloor$ blind spots, where $i,j\in\{0,...,s-1\}$. These mask images are further stacked to form a masked volume $\mathbf{\Omega_{y}}$. 
\item For positions on the denoised volumes $f_{\theta}(\mathbf{\Omega_y})$ corresponding to blind spots, the mask mapper $h(\cdot)$ samples them and assigns them to the same layer. In this way, a globally denoised image $h\left(f_{\theta}\left(\mathbf{{\Omega}_{y}}\right)\right)$ is obtained.
\end{enumerate}
\subsection{Regularized Re-Visible Loss}
Here we present how to use the Blind2Unblind framework for self-supervised denoising. Since blind spot denoising plays a mediating role, the training process should follow the transition from blind to non-blind. However, the pure re-visible loss uses only a single variable that can be backpropagated to optimize both the blind term and the visible term, resulting in unstable training. Therefore, a regular term is introduced to constrain the blind term and stabilize the training procedure. The re-visible loss with regularization is as follows:
\begin{equation}
\label{eq:12}
\begin{split}
\mathcal{L}&=\mathcal{L}_{rev}+\eta\cdot\mathcal{L}_{reg}\\
       &=\Vert h(f_{\theta}(\mathbf{\Omega_{\mathbf{y}}}))+\lambda\hat{f}_{\theta}(\mathbf{y})-(\lambda+1)\mathbf{y}\Vert^2_2\\
&+\eta\cdot\Vert h(f_{\theta}(\mathbf{\Omega_{\mathbf{y}}}))-\mathbf{y}\Vert^2_2.
\end{split}
\end{equation}
where $f_{\theta}(\cdot)$ denotes an arbitrary denoising network with different structures, $h(\cdot)$ is a mask mapper  capable of global modeling, $\eta$ is a fixed hyper-parameter that determines the initial contribution of the blind term and the training stability, and $\lambda$ is a variable hyper-parameter that controls the intensity of visible parts when converting from blind to unblind. The initial and final values of $\lambda$ are $\lambda_{s}$ and $\lambda_{f}$, respectively. To prevent identity mapping, we disable the gradient updating of $\hat{f}_{\theta}(\mathbf{y})$ during training.

\section{Experimental Results}
\label{sec:exp}

\subsection{Implementation Details}
\noindent\textbf{Training Details.} We use the same modified U-Net~\cite{ronneberger2015u} architecture as~\cite{laine2019high,huang2021neighbor2neighbor}. The batch size is 4. We use Adam~\cite{kingma2015adam} as our optimizer, with a weight decay of $1e^{-8}$ to avoid overfitting. The initial learning rate is 0.0003 for synthetic denoising in sRGB space and 0.0001 for real-world denoising, including raw-RGB space and fluorescence microscopy (FM). The learning rate decreases by half every 20 epochs, for 100 training epochs. As for the hyper-parameters included in the re-visible loss, we set $\eta=1, \lambda_{s}=2$ and $\lambda_{f}=20$ empirically. We also randomly crop $128\times128$ patches for training. In practice, masked pixels are a weighted average of pixels in a $3\times3$ neighborhood, as advocated in~\cite{krull2019noise2void}. All models are trained on a server using Python $3.8.5$, Pytorch $1.7.1$~\cite{paszke2019pytorch}, and Nvidia Tesla V100 GPUs.

\noindent\textbf{Datasets for Synthetic Denoising.} Following the setting in~\cite{laine2019high,huang2021neighbor2neighbor}, we select 44,328 images with sizes between $256\times256$ and $512\times512$ pixels from the ILSVRC2012~\cite{deng2009imagenet} validation set as the training set. To obtain reliable average PSNRs, we also repeat the test sets Kodak~\cite{franzen1999kodak}, BSD300~\cite{martin2001database} and Set14~\cite{zeyde2010single} by 10, 3 and 20 times, respectively. Thus, all methods are evaluated with 240, 300, and 280 individual synthetic noise images. Specifically, we consider four types of noise in sRGB space: (1) Gaussian noise with $\sigma=25$, (2) Gaussian noise with $\sigma\in[5,50]$, (3) Poisson noise with $\lambda=30$, (4) Poisson noise with $\lambda\in[5,50]$. Here, values of $\sigma$ are given in 8-bit units. For grayscale image denoising, we use BSD400~\cite{zhang2017beyond} for training and three widely used datasets for testing, including Set12, BSD68~\cite{roth2005fields} and Urban100~\cite{huang2015single}.

\noindent\textbf{Datasets for Real-World Denoising.} Following the setting in~\cite{huang2021neighbor2neighbor}, we take the SIDD~\cite{abdelhamed2018high} dataset collected by five smartphone cameras in 10 scenes under different lighting conditions for real-world denoising in raw-RGB space. We use only raw-RGB images in SIDD Medium Dataset for training and use SIDD Validation and Benchmark Datasets for validation and testing. As for real-world grayscale denoising on FM, we consider FMDD~\cite{zhang2018poisson} dataset, which contains 12 datasets of images captured using either a confocal, two-photon, or widefield microscope. Each dataset contains 20 views with 50 noisy images per view. We select three datasets (Confocal Fish, Confocal Mice and Two-Photon Mice) and train with each dataset, using the 19th view for testing and the rest for training.

\begin{figure*}[ht]
\centering
\setlength{\abovecaptionskip}{0.cm} 
\includegraphics[width=.98\textwidth]{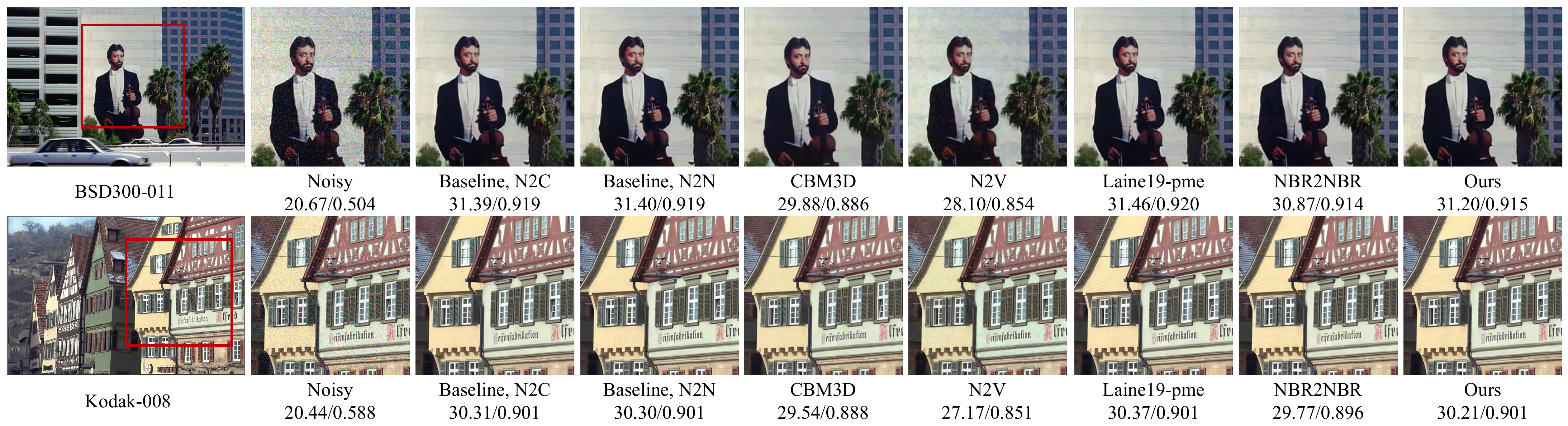}
   \caption{Visual comparison of denoising sRGB images in the setting of $\sigma=25$.}
\label{fig:g25}
\vspace{-0.4cm}
\end{figure*}
\begin{figure*}[ht]
\centering
\setlength{\abovecaptionskip}{0.cm} 
\includegraphics[width=.98\textwidth]{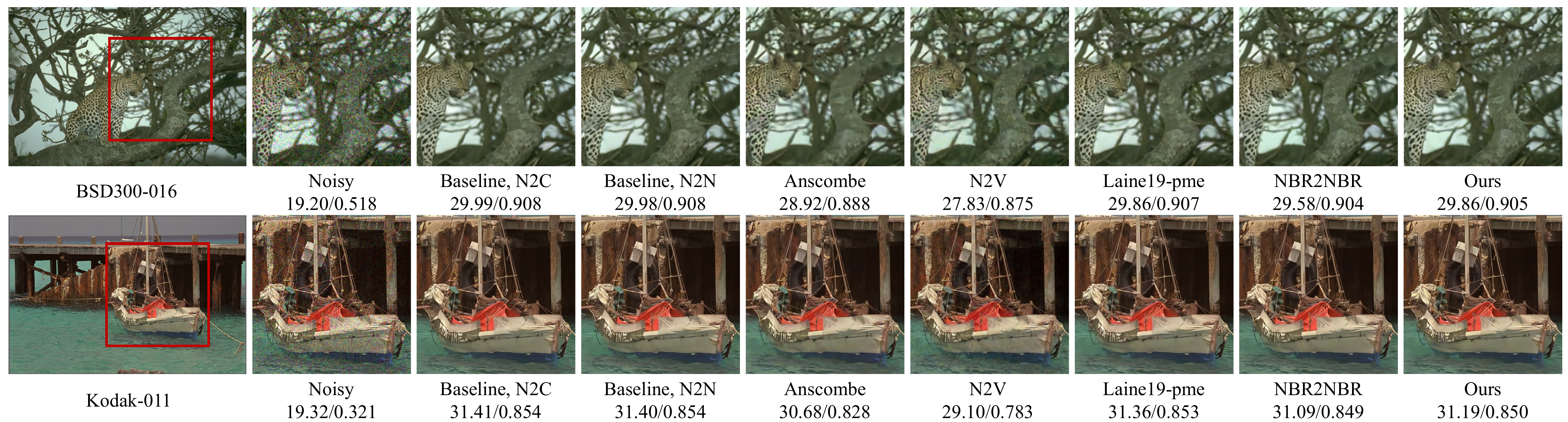}
   \caption{Visual comparison of denoising sRGB images in the setting of $\lambda=30$.}
\label{fig:p30}
\vspace{-0.4cm}
\end{figure*}

\begin{table}[ht]
\scriptsize
  \centering
  \setlength{\abovecaptionskip}{0.1cm} 
  \setlength\tabcolsep{4pt} 
  \begin{tabular}[b]{clccc}
    \toprule
      Noise Type & Method & KODAK & BSD300 & SET14 \\
    \midrule
      \multirow{12}{*}{\shortstack[c]{Gaussian\\$\sigma=25$}} 
        & Baseline, N2C~\cite{ronneberger2015u} & 32.43/0.884 & 31.05/0.879 & 31.40/0.869 \\
        & Baseline, N2N~\cite{lehtinen2018noise2noise} & 32.41/0.884 & 31.04/0.878 & 31.37/0.868 \\
        \cline{2-5}
        & CBM3D~\cite{dabov2007color} & 31.87/0.868 & 30.48/0.861 & 30.88/0.854 \\ 
        & Self2Self~\cite{quan2020self2self} & 31.28/0.864 & 29.86/0.849 & 30.08/0.839 \\
        & N2V~\cite{krull2019noise2void} & 30.32/0.821 & 29.34/0.824 & 28.84/0.802 \\
        & Laine19-mu~\cite{laine2019high} & 30.62/0.840 & 28.62/0.803 & 29.93/0.830 \\
        & Laine19-pme~\cite{laine2019high} & \textbf{32.40/0.883} & \textbf{30.99/0.877} & \textbf{31.36/0.866} \\
        & Noisier2Noise~\cite{moran2020noisier2noise} & 30.70/0.845 & 29.32/0.833 & 29.64/0.832 \\
        & DBSN~\cite{wu2020unpaired} & 31.64/0.856 & 29.80/0.839 & 30.63/0.846 \\
        & R2R~\cite{pang2021recorrupted} & 32.25/\textbf{0.880} & \textbf{30.91/0.872} & \textbf{31.32/0.865} \\
        & NBR2NBR~\cite{huang2021neighbor2neighbor} & 32.08/0.879 & 30.79/\underline{0.873} & 31.09/\underline{0.864} \\
        & Ours & \underline{32.27}/\underline{0.880} & \underline{30.87}/0.872 & \underline{31.27}/\underline{0.864}\\      
    \midrule
      \multirow{12}{*}{\shortstack[c]{Gaussian\\$\sigma\in[5,50]$}} 
        & Baseline, N2C~\cite{ronneberger2015u} & 32.51/0.875 & 31.07/0.866 & 31.41/0.863 \\
        & Baseline, N2N~\cite{lehtinen2018noise2noise} & 32.50/0.875 & 31.07/0.866 & 31.39/0.863 \\
        \cline{2-5}
        & CBM3D~\cite{dabov2007color} & 32.02/0.860 & 30.56/0.847 & 30.94/0.849 \\ 
        & Self2Self~\cite{quan2020self2self} & 31.37/0.860 & 29.87/0.841 & 29.97/0.849 \\
        & N2V~\cite{krull2019noise2void} & 30.44/0.806 & 29.31/0.801 & 29.01/0.792 \\
        & Laine19-mu~\cite{laine2019high} & 30.52/0.833 & 28.43/0.794 & 29.71/0.822 \\
        & Laine19-pme~\cite{laine2019high} & \textbf{32.40}/\underline{0.870} & \textbf{30.95/0.861} & \textbf{31.21}/0.855 \\
        & DBSN~\cite{wu2020unpaired} & 30.38/0.826 & 28.34/0.788 & 29.49/0.814 \\
        & R2R~\cite{pang2021recorrupted} & 31.50/0.850 & 30.56/0.855 & 30.84/0.850 \\
        & NBR2NBR~\cite{huang2021neighbor2neighbor} & 32.10/\underline{0.870} & 30.73/\textbf{0.861} & 31.05/\textbf{0.858} \\
        & Ours & \underline{32.34}/\textbf{0.872} & \underline{30.86}/\textbf{0.861} & \underline{31.14}/\underline{0.857} \\    
    \midrule
      \multirow{12}{*}{\shortstack[c]{Poisson\\$\lambda=30$}} 
        & Baseline, N2C~\cite{ronneberger2015u} & 31.78/0.876 & 30.36/0.868 & 30.57/0.858 \\
        & Baseline, N2N~\cite{lehtinen2018noise2noise} & 31.77/0.876 & 30.35/0.868 & 30.56/0.857 \\
        \cline{2-5}
        & Anscombe~\cite{makitalo2010optimal} & 30.53/0.856 & 29.18/0.842 & 29.44/0.837 \\ 
        & Self2Self~\cite{quan2020self2self} & 30.31/0.857 & 28.93/0.840 & 28.84/0.839 \\
        & N2V~\cite{krull2019noise2void} & 28.90/0.788 & 28.46/0.798 & 27.73/0.774 \\
        & Laine19-mu~\cite{laine2019high} & 30.19/0.833 & 28.25/0.794 & 29.35/0.820 \\
        & Laine19-pme~\cite{laine2019high} & \textbf{31.67/0.874} & \textbf{30.25/0.866} & \textbf{30.47/0.855} \\
        & DBSN~\cite{wu2020unpaired} & 30.07/0.827 & 28.19/0.790 & 29.16/0.814 \\
        & R2R~\cite{pang2021recorrupted} & 30.50/0.801 & 29.47/0.811 & 29.53/0.801 \\
        & NBR2NBR~\cite{huang2021neighbor2neighbor} & 31.44/0.870 & 30.10/\underline{0.863} & 30.29/\underline{0.853} \\
        & Ours & \underline{31.64}/\underline{0.871} & \textbf{30.25}/0.862 & \underline{30.46}/0.852 \\    
    \midrule
      \multirow{12}{*}{\shortstack[c]{Poisson\\$\lambda\in[5,50]$}} 
        & Baseline, N2C~\cite{ronneberger2015u} & 31.19/0.861 & 29.79/0.848 & 30.02/0.842 \\
        & Baseline, N2N~\cite{lehtinen2018noise2noise} & 31.18/0.861 & 29.78/0.848 & 30.02/0.842 \\
        \cline{2-5}
        & Anscombe~\cite{makitalo2010optimal} & 29.40/0.836 & 28.22/0.815 & 28.51/0.817 \\ 
        & Self2Self~\cite{quan2020self2self} & 29.06/0.834 & 28.15/0.817 & 28.83/0.841 \\
        & N2V~\cite{krull2019noise2void} & 28.78/0.758 & 27.92/0.766 & 27.43/0.745 \\
        & Laine19-mu~\cite{laine2019high} & 29.76/0.820 & 27.89/0.778 & 28.94/0.808 \\
        & Laine19-pme~\cite{laine2019high} & \underline{30.88}/0.850 & \underline{29.57}/0.841 & 28.65/0.785 \\
        & DBSN~\cite{wu2020unpaired} & 29.60/0.811 & 27.81/0.771 & 28.72/0.800 \\
        & R2R~\cite{pang2021recorrupted} & 29.14/0.732 & 28.68/0.771 & 28.77/0.765 \\
        & NBR2NBR~\cite{huang2021neighbor2neighbor} & 30.86/\underline{0.855} & 29.54/\underline{0.843} & \underline{29.79}/\underline{0.838} \\
        & Ours & \textbf{31.07/0.857} & \textbf{29.92/0.852} & \textbf{30.10/0.844} \\    
    \bottomrule
  \end{tabular}
  \caption{Quantitative denoising results on synthetic datasets in sRGB space. The highest PSNR(dB)/SSIM among unsupervised denoising methods is highlighted in \textbf{bold}, while the second is \underline{underlined}.}
  \label{tab:sRGB}
  \vspace{-0.4cm}
\end{table}

\noindent\textbf{Details of Experiments.} For fair comparison, we follow the experimental settings of NBR2NBR~\cite{huang2021neighbor2neighbor}. For the baseline, we consider two supervised denoising methods (N2C~\cite{ronneberger2015u} and N2N~\cite{lehtinen2018noise2noise}). We also compare the proposed Blind2Unblind with a traditional approach (BM3D~\cite{dabov2007image}) and seven self-supervised denoising algorithms (Self2Self~\cite{quan2020self2self}, Noise2Void (N2V)~\cite{krull2019noise2void}, Laine19~\cite{laine2019high}, Noisier2Noise~\cite{moran2020noisier2noise}, DBSN~\cite{wu2020unpaired}, R2R~\cite{pang2021recorrupted} and NBR2NBR~\cite{huang2021neighbor2neighbor}). 

In synthetic denoising, we use pre-trained models provided by~\cite{laine2019high} for N2C, N2N, and Laine19, retaining the same network architecture as~\cite{laine2019high,huang2021neighbor2neighbor}. In addition, we use CBM3D~\cite{dabov2007color}, a multi-channel version of BM3D, to perform Gaussian denoising combined with the parameter $\sigma$ estimated by~\cite{chen2015efficient}. For Poisson noise, Anscombe~\cite{makitalo2010optimal} is first run to convert Poisson noise into Gaussian distribution, and then BM3D is used for denoising. For Self2Self, N2V, Noisier2Noise, DBSN, R2R and NBR2NBR, we use the official implementation. 

For real-world denoising in raw-RGB space, all methods use their official implementations and have been retrained on the SIDD Medium Dataset. Note that we split the single-channel raw image into four sub-images according to the Bayer pattern. For BM3D, we denoise the four sub-images individually and then integrate denoised sub-images into a whole image. For deep learning methods, we stack four sub-images to form a four-channel image for denoising and then integrate the denoised image into raw-RGB space. For real-world denoising on FM, we retrain all compared methods using official implementation.

\begin{table}[t]
\scriptsize
  \centering
  \setlength{\abovecaptionskip}{0.1cm} 
  \setlength\tabcolsep{3.5pt} 
  \begin{tabular}[b]{cccccc}
    \toprule
      Noise Type & Method & Network & BSD68 & Set12 & Urban100 \\
    \midrule
      \multirow{3}{*}{\shortstack[c]{Gaussian\\$\sigma=15$}} 
        & N2C~\cite{ronneberger2015u} & U-Net~\cite{ronneberger2015u} & 31.58/0.889 & 32.60/0.899 & 31.95/0.910 \\
      \cline{2-6}
        & R2R~\cite{pang2021recorrupted} & U-Net~\cite{ronneberger2015u} & \textbf{31.54/0.885} & \textbf{32.54/0.897} & \textbf{31.89/0.915} \\
        & Ours & U-Net~\cite{ronneberger2015u} & 31.44/0.884 & 32.46/\textbf{0.897} & 31.79/0.912 \\
    \midrule
      \multirow{3}{*}{\shortstack[c]{Gaussian\\$\sigma=25$}} 
        & N2C~\cite{ronneberger2015u} & U-Net~\cite{ronneberger2015u} & 29.02/0.822 & 30.07/0.852 & 29.01/0.861 \\
      \cline{2-6} 
        & R2R~\cite{pang2021recorrupted} & U-Net~\cite{ronneberger2015u} & \textbf{28.99}/0.818 & 30.06/0.851 & \textbf{29.17/0.865} \\
        & Ours & U-Net~\cite{ronneberger2015u} & \textbf{28.99/0.820} & \textbf{30.09/0.854} & 29.05/0.864 \\
    \midrule
      \multirow{3}{*}{\shortstack[c]{Gaussian\\$\sigma=50$}} 
        & N2C~\cite{ronneberger2015u} & U-Net~\cite{ronneberger2015u} & 26.08/0.715 & 26.88/0.777 & 25.35/0.766 \\
      \cline{2-6} 
        & R2R~\cite{pang2021recorrupted} & U-Net~\cite{ronneberger2015u} & 26.02/0.705 & 26.86/0.771 & 25.48/0.768 \\
        & Ours & U-Net~\cite{ronneberger2015u} & \textbf{26.09/0.715} & \textbf{26.91/0.776} & \textbf{25.57/0.868} \\
    \bottomrule
  \end{tabular}
  \caption{Grayscale image denoising results.}
  \label{tab:gray}
  \vspace{-0.7cm}
\end{table}

\subsection{Results for Synthetic Denoising}
The quantitative comparison results of synthetic denoising for Gaussian and Poisson noise can be seen in~\Cref{tab:sRGB}. For Gaussian noise, whether the noise level is fixed or variable, our approach significantly outperforms the traditional denoising method BM3D and five self-supervised denoising methods, including Self2Self, N2V, Laine19-mu, DBSN, and NBR2NBR. R2R also performs worse than our method under variable Gaussian noise, despite a known noise prior. For fixed Poisson noise, our approach shows comparable performance to Laine19-pme, which is based on explicit noise modeling. However, explicit noise modeling means strong prior, leading to poor performance on real data. The following experiments on real-world datasets also illustrate this problem. When the Poisson noise level is variable, our method outperforms almost all methods to be compared, including Laine19-pme. Moreover, our approach outperforms the supervision baseline (N2C) with a gain of 0.13 dB and 0.08 dB on BSD300 and SET14, respectively.  Namely, our method shows more superior performance for removing Poisson noise. In addition, whether the noise is fixed or variable, our method significantly outperforms the SOTA method NBR2NBR, with a maximum gain of 0.38 dB.~\Cref{fig:g25,fig:p30} illustrate the sRGB denoising results in the setting of $\sigma=25$ and $\lambda=30$, respectively. Compared with the recent best NBR2NBR, our method recovers more texture details.~\Cref{tab:gray} conducts additional experiments on grayscale images. As the noise increases, our method gradually outperforms R2R, because re-corruption loses more information, offsetting the noise prior advantage. Moreover, our method infers directly, while R2R repeats 50 times.
\begin{figure}[ht]
\centering
\setlength{\abovecaptionskip}{0.cm} 
\includegraphics[width=.42\textwidth]{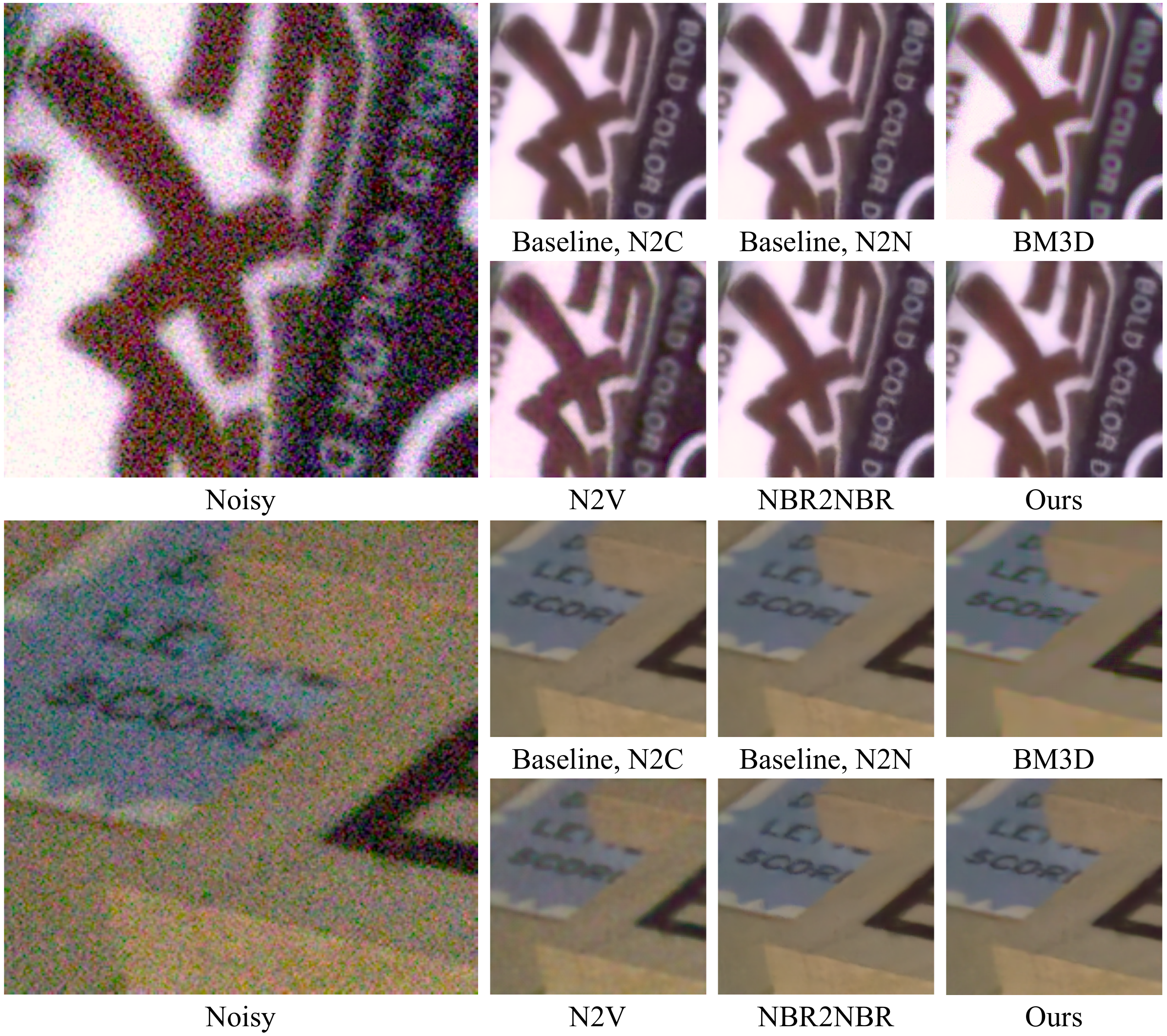}
   \caption{Visual comparison of denoising raw-RGB images in the challenging SIDD benchmark. The conversion from raw-RGB to sRGB is performed using the official ISP tool.}
\label{fig:raw}
\end{figure}
\begin{table}[ht]
\scriptsize
  \centering
  \vspace{-0.2cm}
  \setlength{\abovecaptionskip}{0.1cm} 
  \setlength\tabcolsep{6pt}
  \begin{tabular}[b]{llcc}
    \toprule
      \multirow{2}{*}{Methods} 
      & \multirow{2}{*}{Network} 
      & SIDD & SIDD\\
      & & Benchmark & Validation\\
    \midrule
      Baseline, N2C~\cite{ronneberger2015u} & U-Net~\cite{ronneberger2015u} & 50.60/0.991 & 51.19/0.991 \\
      Baseline, N2N~\cite{lehtinen2018noise2noise} & U-Net~\cite{ronneberger2015u} & 50.62/0.991 & 51.21/0.991 \\
    \midrule
      BM3D~\cite{dabov2007image} & - & 48.60/0.986 & 48.92/0.986 \\
      N2V~\cite{krull2019noise2void} & U-Net~\cite{ronneberger2015u} & 48.01/0.983 & 48.55/0.984 \\
      Laine19-mu (Gaussian)~\cite{laine2019high} & U-Net~\cite{ronneberger2015u} & 49.82/0.989 & 50.44/0.990 \\
      Laine19-pme (Gaussian)~\cite{laine2019high} & U-Net~\cite{ronneberger2015u} & 42.17/0.935 & 42.87/0.939 \\
      Laine19-mu (Poisson)~\cite{laine2019high} & U-Net~\cite{ronneberger2015u} & 50.28/0.989 & 50.89/0.990 \\
      Laine19-pme (Poisson)~\cite{laine2019high} & U-Net~\cite{ronneberger2015u} & 48.46/0.984 & 48.98/0.985 \\
      DBSN~\cite{wu2020unpaired} & DBSN~\cite{wu2020unpaired} & 49.56/0.987 & 50.13/0.988 \\
      R2R~\cite{pang2021recorrupted} & U-Net~\cite{ronneberger2015u} & 46.70/0.978 & 47.20/0.980 \\
      NBR2NBR~\cite{huang2021neighbor2neighbor} & U-Net~\cite{ronneberger2015u} & 50.47/0.990 & 51.06/0.991 \\
      Ours & U-Net~\cite{ronneberger2015u} & \textbf{50.79/0.991} & \textbf{51.36/0.992} \\
    \bottomrule
  \end{tabular}
  \caption{Quantitative denoising results on SIDD benchmark and validation datasets in raw-RGB space.}
  \label{tab:rawRGB}
  \vspace{-0.7cm}
\end{table}

\subsection{Results for raw-RGB Denoising}
In raw-RGB space,~\Cref{tab:rawRGB} shows the quality scores for quantitative comparisons on SIDD Benchmark and SIDD Validation. Note that the online website~\cite{web:siddbenchmark} evaluates the quality scores for the SIDD Benchmark. In general, the proposed method outperforms the state-of-the-art (NBR2NBR) by 0.32 dB and 0.20 dB for the benchmark and validation. Because the transition from blind to non-blind preserves all information contained within noisy images and mitigates the effects of over smoothing among neighboring pixels. Moreover, our method performs far better than R2R in the real world. Two primary reasons are lossy re-corruption and inaccurate noise priors in R2R. Interestingly, our approach outperforms the baseline (N2C) by 0.19 dB and 0.17 dB in the benchmark and validation, indicating improved generalization. The raw-RGB denoising performance in the real world demonstrates that the proposed approach is competitive in the presence of complex noise patterns.~\Cref{fig:raw} demonstrates our approach's superiority over alternative methods. Our method recovers more texture details and has a higher degree of continuity, whereas NBR2NBR exhibits obvious sawtooth diffusion defects due to the training scheme of approximating neighboring pixels.

\subsection{Results for FM Denoising.}
\begin{figure}[h]
\centering
\setlength{\abovecaptionskip}{0.cm} 
\includegraphics[width=.42\textwidth]{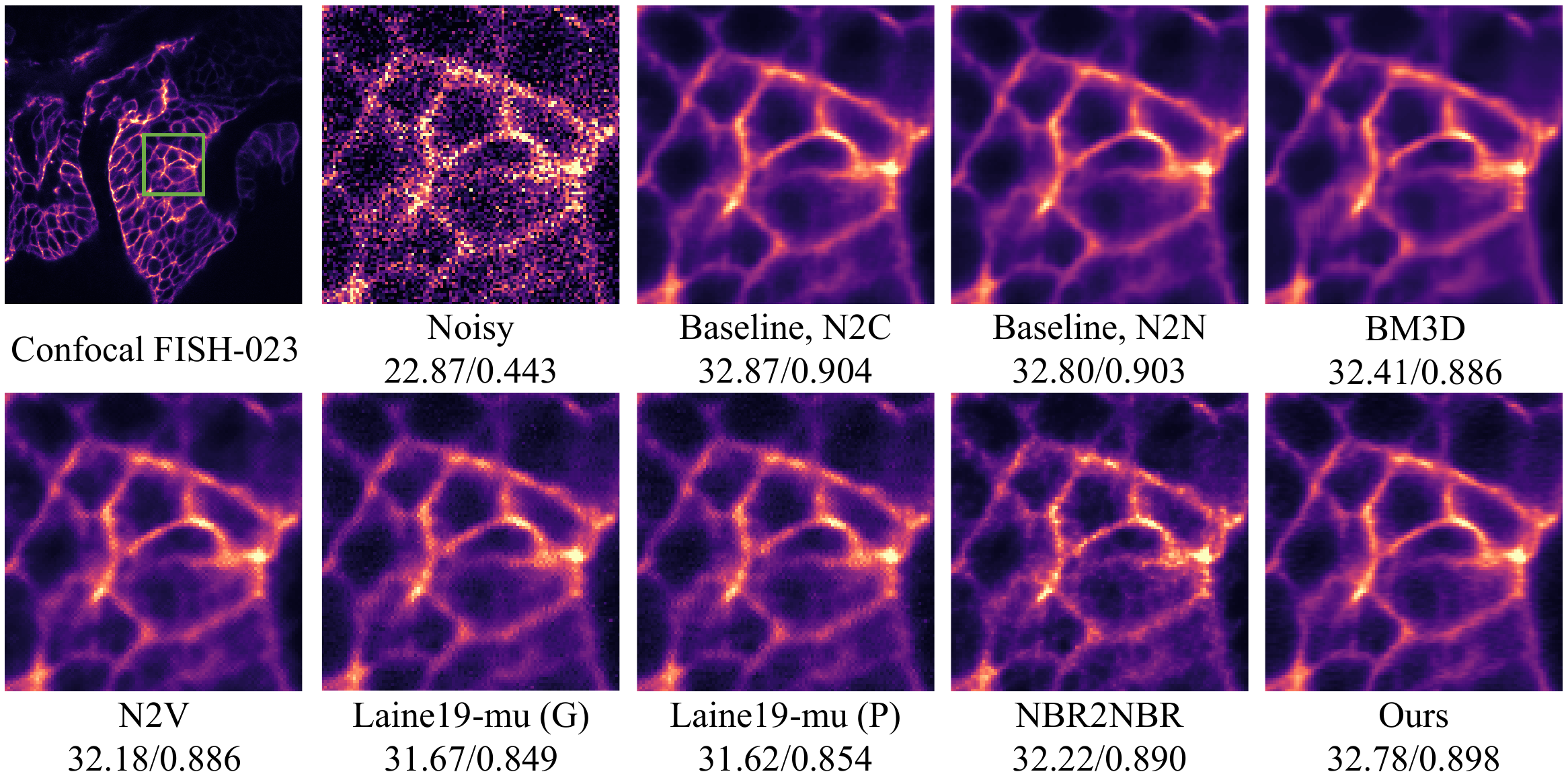}
   \caption{Visual comparison of denoising FM images.}
\label{fig:fm}
\vspace{-0.6cm}
\end{figure}

\Cref{tab:fm} shows the quantitative comparison of real-world fluorescence microscopy datasets. Our approach shows competitive performance against other methods. In particular, our method greatly outperforms self-supervised methods and  slightly outperforms supervised methods (N2C and N2N) on both Confocal Mice and Two-Photon Mice. The results further confirm the superior performance of our method on complex real-world noise patterns.~\Cref{fig:fm} shows a visual comparison of denoising FM images.

\begin{table}[ht]
\scriptsize
  \centering
  \setlength{\abovecaptionskip}{0.1cm} 
  \setlength\tabcolsep{4pt}
  \begin{tabular}[b]{@{}llccc@{}}
    \toprule
      \multirow{2}{*}{Methods} 
      & \multirow{2}{*}{Network} 
      & Confocal & Confocal & Two-Photon\\
      & & Fish & Mice & Mice\\
    \midrule
      Baseline, N2C~\cite{ronneberger2015u} & U-Net~\cite{ronneberger2015u} & 32.79/0.905 & 38.40/0.966 & 34.02/0.925 \\
      Baseline, N2N~\cite{lehtinen2018noise2noise} & U-Net~\cite{ronneberger2015u} & 32.75/0.903 & 38.37/0.965 & 33.80/0.923 \\
    \midrule
      BM3D~\cite{dabov2007image} & - & 32.16/0.886 & 37.93/0.963 & 33.83/\textbf{0.924} \\
      N2V~\cite{krull2019noise2void} & U-Net~\cite{ronneberger2015u} & 32.08/0.886 & 37.49/0.960 & 33.38/0.916 \\
      Laine19-mu (G)~\cite{laine2019high} & U-Net~\cite{ronneberger2015u} & 31.62/0.849 & 37.54/0.959 & 32.91/0.903 \\
      Laine19-pme (G)~\cite{laine2019high} & U-Net~\cite{ronneberger2015u} & 23.30/0.527 & 31.64/0.881 & 25.87/0.418 \\
      Laine19-mu (P)~\cite{laine2019high} & U-Net~\cite{ronneberger2015u} & 31.59/0.854 & 37.30/0.956 & 33.09/0.907 \\
      Laine19-pme (P)~\cite{laine2019high} & U-Net~\cite{ronneberger2015u} & 25.16/0.597 & 37.82/0.959 & 31.80/0.820 \\
      NBR2NBR~\cite{huang2021neighbor2neighbor} & U-Net~\cite{ronneberger2015u} & 32.11/0.890 & 37.07/0.960 & 33.40/0.921 \\
      Ours & U-Net~\cite{ronneberger2015u} & \textbf{32.74/0.897} & \textbf{38.44/0.964} & \textbf{34.03}/0.916 \\
    \bottomrule
  \end{tabular}
  \caption{Quantitative denoising results on Confocal Fish, Confocal Mice and Two-Photon Mice. G is for Gaussian and P is for Poisson.}
  \label{tab:fm}
  \vspace{-0.4cm}
\end{table}

\subsection{Ablation Study}
This section conducts ablation studies on the loss function, mask strategy, visible term, and regular term. Note that PSNR(dB)/SSIM is evaluated on Kodak.

\noindent\textbf{Ablation study on Loss Function.}~\Cref{tab:loss} performs ablation experiments under two loss conditions. Since $\hat{f}_{\theta}(\mathbf{y})$ in~\cref{eq:7} is suppressed by the mask term $h(f_{\theta}(\mathbf{\Omega_{\mathbf{y}}}))$, the performance of $\mathcal{L}_{B}$ is much lower than that of $\mathcal{L}_{A}$.

\noindent\textbf{Ablation study on Mask Mapper.} To evaluate our global-aware mask mapper (GM), we compare it with the random mask (RM) as advocated in Noise2Void~\cite{krull2019noise2void}. Details are provided in the supplementary
materials.~\Cref{tab:mask} lists the performance of different mask strategies in self-supervised denoising. We can see that our GM considerably outperforms RM for all four noise patterns. Moreover, GM+V performs much better combined with re-visible loss, while RM+V is severely suppressed and cannot converge.

\noindent\textbf{Ablation study on Visible Term.} The hyper-parameter $\lambda_{f}$ is positively correlated with the visibility of the non-blind term.~\Cref{tab:visible} shows that the weight of the visible item is not the larger, the better: 1) When $\lambda_{f}=20$, our approach achieves or approaches the best performance on four noise patterns. 2) When $\lambda_{f}$ varies from 20 to 40, the performance on Poisson noise decreases. In contrast, the performance on Gaussian noise remains fixed. 3) When $\lambda_{f}>40$, the performance is almost unchanged. In this paper, we set $\lambda_{f}=20$.

\begin{table}[t]
\scriptsize
  \centering
  \setlength{\abovecaptionskip}{0.1cm} 
  \setlength\tabcolsep{7pt}
  \begin{tabular}[b]{@{}ccccc}
    \toprule
      Loss Type & $\sigma=25$ & $\sigma\in[5,50]$ & $\lambda=30$ & $\lambda\in[5,50]$ \\
    \midrule
      $\mathcal{L}_{A}$ & \textbf{32.27/0.880} & \textbf{32.34/0.872} & \textbf{31.64/0.871} & \textbf{31.07/0.857} \\
      $\mathcal{L}_{B}$ & 31.37/0.873 & 30.85/0.851 & 30.95/0.863 & 30.01/0.840 \\
    \bottomrule
  \end{tabular}
  \caption{Ablation study on the loss function. $\mathcal{L}_{A}$ and $\mathcal{L}_{B}$ denote~\cref{eq:6} and~\cref{eq:7}.}
  \label{tab:loss}
  \vspace{-0.3cm}
\end{table}
\begin{table}[t]
\scriptsize
  \centering
  \setlength{\abovecaptionskip}{0.1cm} 
  \setlength\tabcolsep{5pt}
  \begin{tabular}[b]{@{}lcccc}
    \toprule
      Noise Type & RM & GM & RM+V & GM+V \\
    \midrule
      Gaussian $\sigma=25$ & 30.32/0.831 & 30.56/0.839 & - / - & \textbf{32.27/0.880} \\
      Gaussian $\sigma\in[5,50]$ & 30.23/0.822 & 30.46/0.831 & - / - & \textbf{32.34/0.872} \\
      Poisson $\lambda=30$ & 29.89/0.821 & 30.11/0.830 & - / - & \textbf{31.64/0.871} \\
      Poisson $\lambda\in[5,50]$ & 29.26/0.796 & 29.67/0.817 & - / - & \textbf{31.07/0.857} \\
    \bottomrule
  \end{tabular}
  \caption{Ablation study on mask mappers. RM, GM and V denote random mask mapper, global-aware mask mapper, and re-visible loss. - / - means cannot converge.}
  \label{tab:mask}
  \vspace{-0.3cm}
\end{table}
\begin{table}[t]
\scriptsize
  \centering
  \setlength{\abovecaptionskip}{0.1cm} 
  \setlength\tabcolsep{3.5pt}
  \begin{tabular}[b]{@{}lcccccc@{}}
    \toprule
      Noise Type & $\lambda_{f}=2$ & $\lambda_{f}=20$ & $\lambda_{f}=40$ & $\lambda_{f}=100$ \\
    \midrule
      Gaussian $\sigma=25$ & 32.11/0.879 & \textbf{32.27/0.880} & \textbf{32.27}/0.879 & 32.26/0.879 \\
      Gaussian $\sigma\in[5,50]$ & 32.00/0.871 & 32.34/0.872 & 32.34/0.872 & \textbf{32.35/0.873} \\
      Poisson $\lambda=30$ & 31.47/0.869 & \textbf{31.64/0.871} & 31.52/0.869 & 31.51/0.869  \\
      Poisson $\lambda\in[5,50]$ & 30.94/0.854 & \textbf{31.07/0.857} & 31.02/0.855 & 31.01/0.854  \\
    \bottomrule
  \end{tabular}
  \caption{Ablation study on visible term. Note that $\eta=1, \lambda_{s}=2$.}
  \label{tab:visible}
  \vspace{-0.3cm}
\end{table}
\begin{table}[t]
\scriptsize
  \centering
  \setlength{\abovecaptionskip}{0.1cm} 
  \setlength\tabcolsep{3.5pt}
  \begin{tabular}[b]{@{}lcccc@{}}
    \toprule
      Noise Type & $\eta=0$ & $\eta=1$ & $\eta=2$ & $\eta=3$ \\
    \midrule
      Gaussian $\sigma=25$ & 32.19/0.877 & 32.27/0.880 & \textbf{32.32/0.881} & 32.21/0.878 \\
      Gaussian $\sigma\in[5,50]$ & 32.21/0.870 & \textbf{32.34/0.872} & 32.31/\textbf{0.872} & 32.23/0.871 \\
      Poisson $\lambda=30$ & 31.54/0.869 & \textbf{31.64/0.871} & 31.54/0.869 & 31.52/0.869 \\
      Poisson $\lambda\in[5,50]$ & 31.03/0.856 & \textbf{31.07/0.857} & 31.06/\textbf{0.857} & 31.04/\textbf{0.857} \\
    \bottomrule
  \end{tabular}
  \caption{Ablation study on regular term. Here, $\lambda_{s}=2, \lambda_{f}=20$.}
  \label{tab:reg}
  \vspace{-0.6cm}
\end{table}

\noindent\textbf{Ablation study on Regularization Term.} The hyper-parameter $\eta$ controls the stability of re-visible loss.~\Cref{tab:reg} shows that the denoising performance first increases and then decreases on four noise patterns. In particular, when $\eta=1$, the performance is at or closed to the best, which proves that the regular term strengthens the training stability and improves the performance. In this paper, we set $\eta=1$.

\section{Conclusion}
\label{sec:conc}
We propose Blind2Unblind, a novel self-supervised denoising framework, which achieves lossless denoising through the transition from blind to non-blind and enables blindspot schemes to use complete noisy images without information loss. The global-aware mask mapper samples the denoised volume at blind spots, and then re-visible loss realizes non-blind denoising under visible blind spots. Extensive experiments have shown the superiority of our approach against compared methods, especially for complex noise patterns.

\section*{Acknowledgments}
This work is jointly supported by National Natural Science Foundation of China (32171461), the Strategic Priority Research Program of Chinese Academy of Science (XDA16021104, XDB32030208), Program of Beijing Municipal Science \& Technology Commission (Z201100008420004), and Science and Technology Innovation 2030 Major Projects of China (2021ZD0204503).

\clearpage

{\small
\bibliographystyle{ieee_fullname}
\bibliography{egbib}
}

\clearpage

\appendix
\section{Training Framework for Blind2Unblind}
The training framework for Blind2Unblind is shown in~\cref{ag:b2u}.
\vspace{-0.3cm}
\begin{algorithm}[htp]
\SetAlgoLined
\caption{Blind2Unblind}
\label{ag:b2u}
\KwIn{A set of noisy images $Y=\{\mathbf{y}_i\}^{n}_{i=1}$;\\
		\qquad\quad\ \!Denoising network $f_{\theta}(\cdot)$;\\
		\qquad\quad\ \!Hyper-parameters $\eta,\lambda$;}

\While{not converged}
{Sample a noisy image $\mathbf{y}$\;
 Generate a global masker $\mathbf{\Omega}_{(\cdot)}$\;
 Derive a masked volume $\mathbf{\Omega_{y}}$, where $\mathbf{\Omega_{y}}$ is the network input, and $\mathbf{y}$ is the network target\;
 For the network input $\mathbf{\Omega_{y}}$, derive the denoised volume $f_{\theta}(\mathbf{\Omega_{y}})$\;
 Global mask mapper $h_{(\cdot)}$ samples the denoised volume $f_{\theta}(\mathbf{\Omega_{y}})$ at blind spots, then obtain a blind denoised image $h(f_{\theta}(\mathbf{\Omega_{y}}))$\;
 For the original noisy image $\mathbf{y}$, derive the visible denoised image $\hat{f}_{\theta}(\mathbf{y})$ without gradients\;
 Calculate re-visible loss $\mathcal{L}_{rev}=\Vert h(f_{\theta}(\mathbf{\Omega_{\mathbf{y}}}))+\lambda\hat{f}_{\theta}(\mathbf{y})-(\lambda+1)\mathbf{y}\Vert^2_2$\;
 Calculate regularization $\mathcal{L}_{reg}=\Vert h(f_{\theta}(\mathbf{\Omega_{\mathbf{y}}}))-\mathbf{y}\Vert^2_2$\;
 Update network parameters $\theta$ by minimizing the regularized re-visible loss $\mathcal{L}_{rev}+\eta\cdot\mathcal{L}_{reg}$\;}
\end{algorithm}
\vspace{-0.6cm}
\section{Details of Interpolation from Neighbors}
\label{sec:intp}
\begin{figure}[htp]
\centering
\setlength{\abovecaptionskip}{0.cm} 
\includegraphics[width=.48\textwidth]{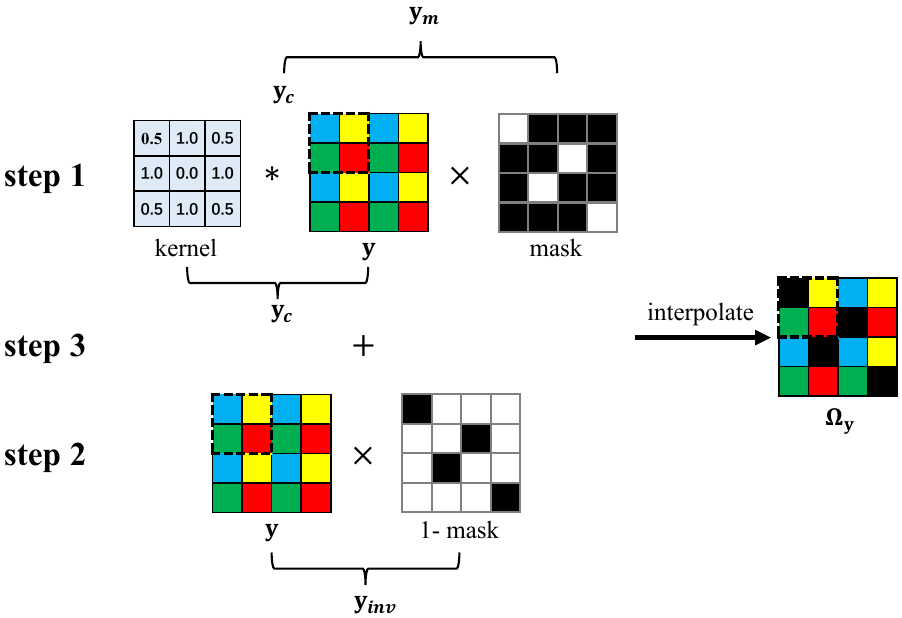}
   \caption{Details of interpolation from neighbors.}
\label{fig:intp}
\vspace{-0.5cm}
\end{figure}

\Cref{fig:intp} shows the workflow of interpolation from neighbors. The workflow can be divided into the following three steps: 1) The mask is generated by random masking each $2\times2$ cells in image $\mathbf{y}$. The kernel convolves image $\mathbf{y}$ with stride $1$ and padding $1$ to produce $\mathbf{y}_c$. Then, $\mathbf{y}_m$ is obtained via Hadamard product $\mathbf{y}_c\circ mask$. 2) We perform Hadamard product $\mathbf{y} \circ (1-mask)$ to gernerate $\mathbf{y}_{inv}$. 3) Sum by $\mathbf{y}_m$ and $\mathbf{y}_{inv}$, we finally obtain the masked image $\mathbf{\Omega_y}$.
\section{Details of Random Mask Strategy}
\label{sec:rm}
\begin{figure}[htp]
\centering
\setlength{\abovecaptionskip}{0.cm} 
\includegraphics[width=.48\textwidth]{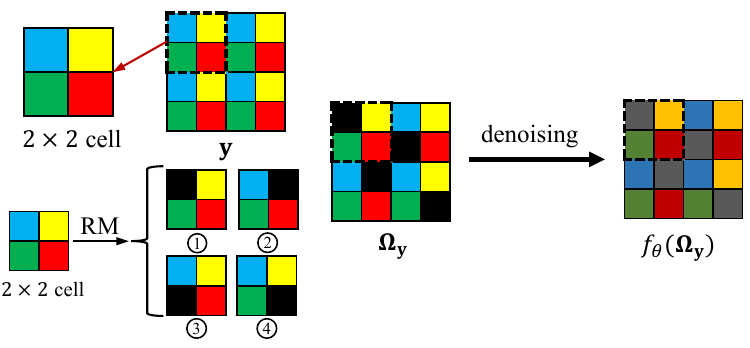}
   \caption{Details of random mask strategy.}
\label{fig:rms}
\vspace{-0.6cm}
\end{figure}

The illustration of the random mask strategy is presented in~\Cref{fig:rms}. The image y is divided into several blocks with 2x2 cells. A specific pixel in each cell is randomly set as a blind spot. Namely, there are four ways for random masking of 2x2 cells. After random masking, the masked image $\mathbf{\Omega_{y}}$ is fed into the denoising network to generate the denoised image $f_{\theta}(\mathbf{\Omega_{y}})$.
\section{More Experimental Results}
\begin{figure*}[htp]
\centering
\setlength{\abovecaptionskip}{0.cm} 
\includegraphics[width=\textwidth]{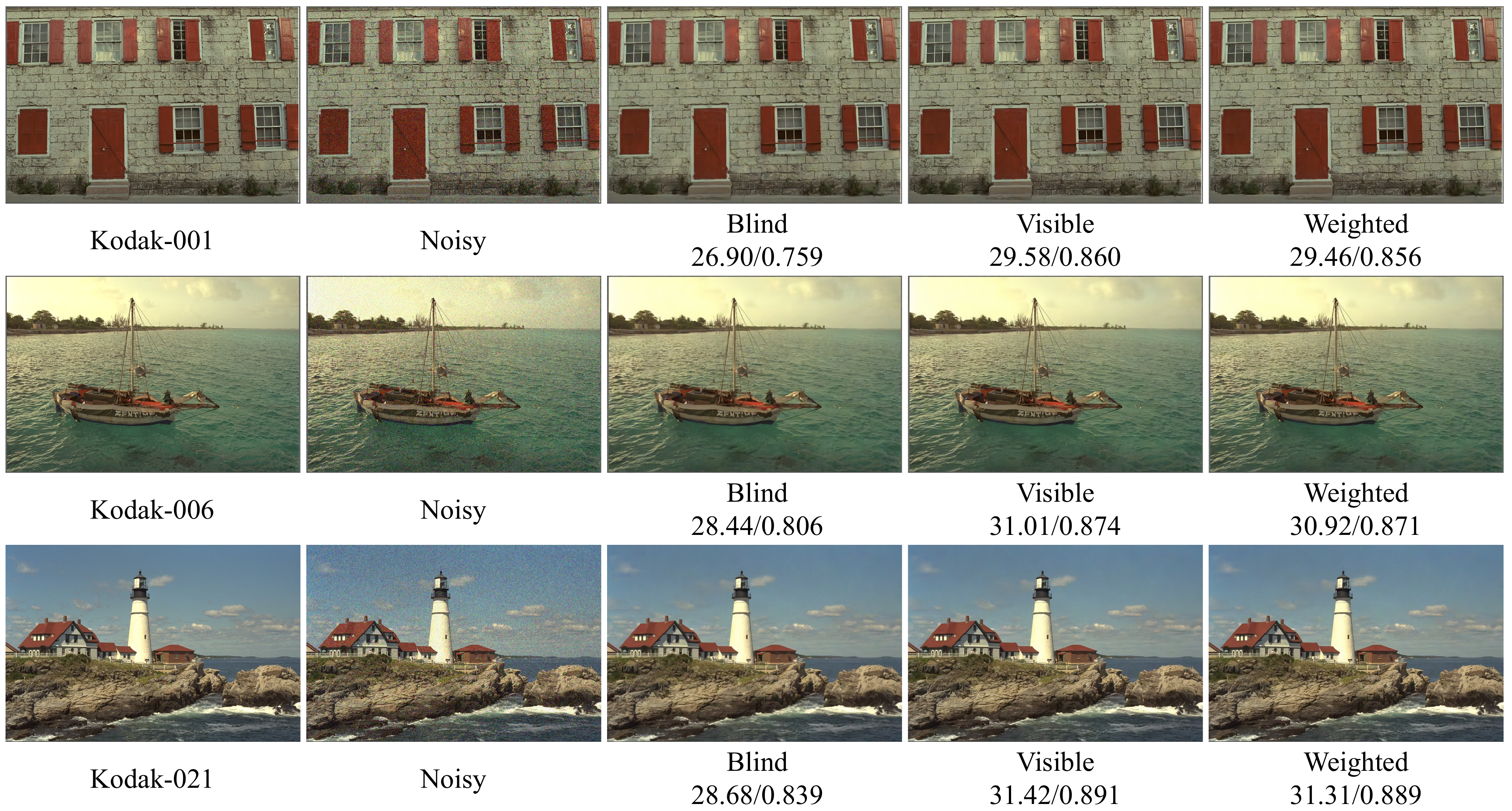}
   \caption{The steps of our proposed method while denoising sRGB images in the
setting of $\sigma=25$. Blind denotes $h(f_{\theta}(\mathbf{\Omega_{\mathbf{y}}}))$, Visible denotes $\hat{f}_{\theta}(\mathbf{y})$, and Weighted denotes $\frac{h(f^{*}_{\theta}(\mathbf{\Omega_{\mathbf{y}}}))+\lambda\hat{f}^{*}_{\theta}(\mathbf{y})}{\lambda+1}$.}
\label{fig:steps}
\vspace{-0.1cm}
\end{figure*}

\begin{figure*}[ht]
\centering
\setlength{\abovecaptionskip}{0.cm} 
\includegraphics[width=\textwidth]{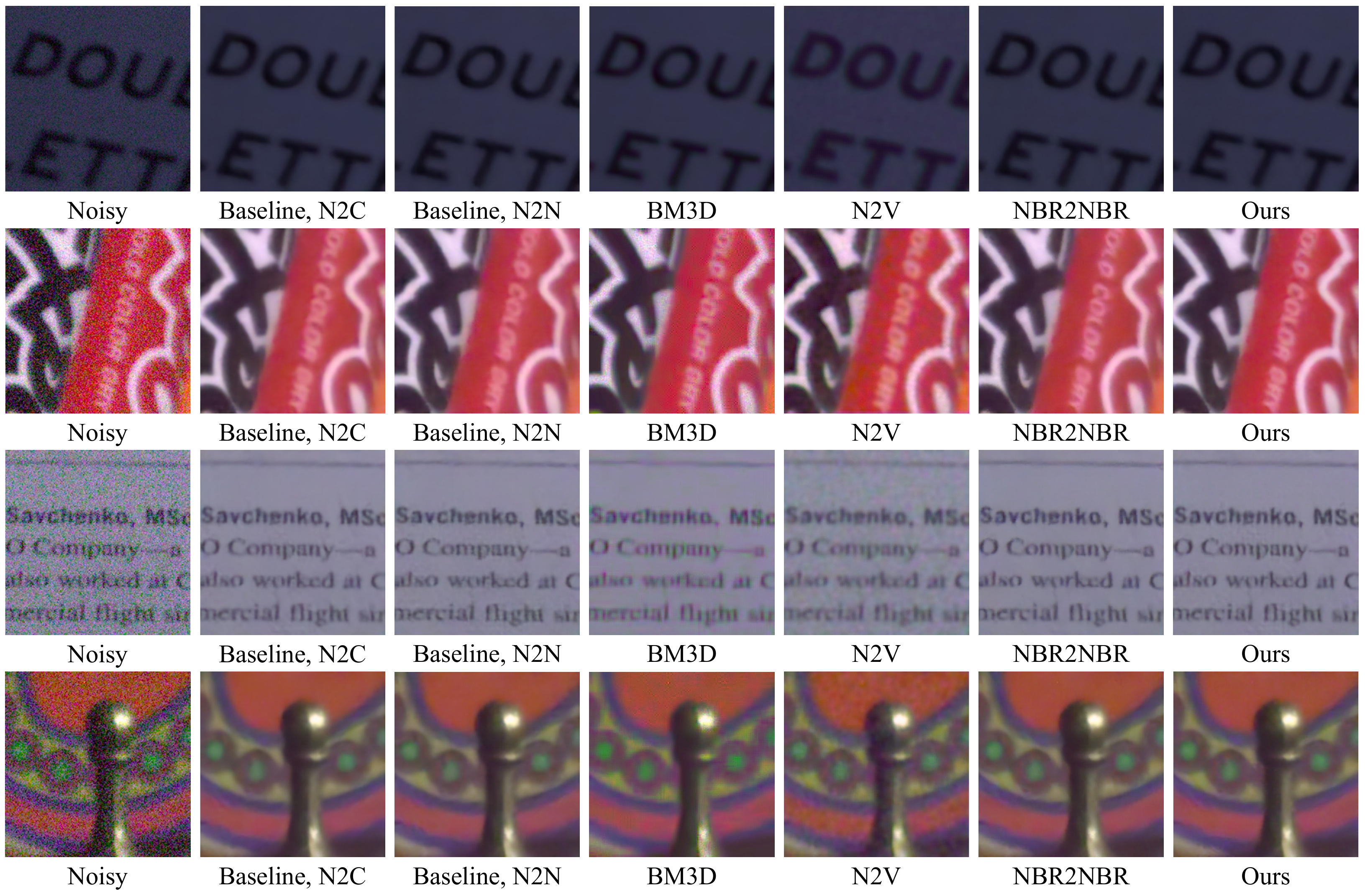}
   \caption{Visual comparison of denoising raw-RGB images in the
challenging SIDD benchmark.}
\label{fig:rawRGB2}
\vspace{-0.4cm}
\end{figure*}

\Cref{fig:steps} illustrates the steps of our proposed method while denoising sRGB images in the
setting of $\sigma=25$.~\Cref{fig:rawRGB2} shows the visual comparison of denoising raw-RGB images in the
challenging SIDD benchmark.

\end{document}